\documentclass[%
 aip,
 amsmath,amssymb,
reprint,%
]{revtex4-2}

\usepackage{graphicx}% Include figure files
\usepackage{dcolumn}% Align table columns on decimal point
\usepackage{bm}% bold math
\usepackage[utf8]{inputenc}
\usepackage[T1]{fontenc}
\usepackage{mathptmx}
\usepackage{etoolbox}

\makeatletter
\def\@email#1#2{%
 \endgroup
 \patchcmd{\titleblock@produce}
  {\frontmatter@RRAPformat}
  {\frontmatter@RRAPformat{\produce@RRAP{*#1\href{mailto:#2}{#2}}}\frontmatter@RRAPformat}
  {}{}
}%
\makeatother
\begin{document}

\preprint{AIP/123-QED}

\title[Extreme soliton collisions]{Statistical properties of extreme soliton collisions}
% Force line breaks with \\
\author{A.V. Slunyaev}
\author{T.V. Tarasova}%
 \email{slunyaev@ipfran.ru}
\affiliation{ National Research University-Higher School of Economics, 25 Bol’shaya Pechorskaya Street, Nizhny Novgorod 603950, Russia \\
and Institute of Applied Physics RAS, 46 Ulyanova Street, Nizhny Novgorod 603950, Russia%\\This line break forced with \textbackslash\textbackslash
}%

% \homepage{http://www.Second.institution.edu/~Charlie.Author.}
%\affiliation{%
%Second institution and/or address%\\This line break forced% with \\
%}%

\date{\today}% It is always \today, today,

\begin{abstract}
Synchronous collisions between a large number of solitons are considered in the context of a statistical description. It is shown that during the interaction of solitons of the same signs the wave field is effectively smoothed out. When the number of solitons increases and the sequence of their amplitudes decay slower, the focused wave becomes even smoother and the statistical moments get frozen for a long time. This quasi-stationary state is characterized by greatly reduced statistical moments and by the density of solitons close to some critical value. This state may be treated as the small-dispersion limit, what makes it possible to analytically estimate all high-order  statistical moments. While the focus of the study is made on the Korteweg--de Vries equation and its modified version, a much broader applicability of the results to equations that support soliton-type solutions is discussed.
\end{abstract}

\maketitle

\begin{quotation}
	In this communication we show how synchronous collisions between a great number of solitons realize the states with the maximum soliton density which corresponds to the minimal allowed spatial size of the soliton ensemble. Within a broad class of evolution equations these particular states of solitons of the same sign are characterized by new approximate conserved quantities in the form of statistical moments. This happens due to the effective smoothing of the wave field stemming from the great ratio of the nonlinearity vs dispersion parameter. These states correspond to the minimum values of statistical moments for interacting unipolar solitons, which are evaluated analytically.
\end{quotation}

%=======================================================
%\emph{Introduction.---}
\section{Introduction} \label{sec:Intro}
%=======================================================
%
The soliton turbulence is a challenging topic of recent research which is highly interesting from both, the mathematical and the physical points of view. In physics it describes strongly coherent wave states when the assumptions of a wave (weak) turbulence are violated, and thus the quasi-Gaussian probabilistic description completely fails. In mathematics, it represents deterministic chaos-like behaving systems which are formally solvable by means of the Inverse Scattering Transform, but its machinery is often too much complicated to write down explicit solutions or relations.

The kinetic theory for a soliton gas \cite{Zakharov1971,ElKamchatnov2005} describes the transport of the soliton spectral density. Due to the failure of a linear superposition property, the kinetic theory cannot help to compute wave fields as such. In particular, the occurrence and probability of extreme events cannot be evaluated. 
It was shown in Refs.~\onlinecite{SlunyaevPelinovsky2016,Slunyaev2019} within the Korteweg--de Vries-type equations, that interactions of solitons of the same sign lead to some reduction of the wave field amplitude, whereas collisions of solitons of different signs (which can co-exist in the modified Korteweg -- de Vries or the Gardner equations of the focusing types) can produce very high waves (so-called rogue waves).
%Depending on the sign of solitons within the Korteweg--de Vries-type equations, their interactions can lead to either generation of very high waves (so-called rogue waves) or to reduction of the wave field amplitudes \cite{SlunyaevPelinovsky2016,Slunyaev2019}. 
%The latter case seems to be generic for solitons of the same sign, hence one may conclude that collisions of unipolar solitons are not prone to generating extreme events. 
Meanwhile, the question of a meaningful description of the amplitude probability distribution in essentially coherent wave ensembles remains unclear.

It was pointed out recently \cite{El2016,PelinovskyShurgalina2017}, that the standard definition of the variance
\begin{align}\label{Variance}
	\sigma^2 = \overline{ ( u - \overline{u} ) ^2} = 
	\overline{u^2} - \overline{u}^2 > 0
\end{align}
for a sign-defined wave field $u(x) \ge 0$ imposes a formal limit on the quantity which has the meaning of the soliton density, $\rho = N/L > 0$, $\rho \le \rho_{cr}$, where $N$ is the number of solitons per the spatial interval $L$. 
Here the overline means averaging over the interval $L$, which may be converted to the averaging in the spectral plane of the associated scattering problem to the given integrable equation when $L$ tends to infinity, see Ref.~\onlinecite{El2016}.
The distribution of eigenvalues may provide information on a chance of large-amplitude wave generation \cite{Soto-Crespoetal2016}.  
%In this work we will focus on the extreme states when solitons of a same sign are located very densily.

%In the particular example of the Korteweg -- de Vries equation (KdV)
Let us consider an example of the Korteweg -- de Vries (KdV) equation
\begin{align}\label{KdV}
	u_t + 6 u u_{x} + u_{xxx} = 0.
\end{align}
Its exact $N$-soliton solution $u_N(x,t)$ can be obtained via consecutive Darboux transformations \cite{MatveevSalle1991} which allow a compact representation
\begin{align}\label{NKdVSoliton}
	u_N(x,t) = 2\frac{\partial^2}{\partial x^2} \ln{W(\psi_{1},\psi_{2},...,\psi_{N})}.
\end{align}
Here $W(\cdot)$ denotes the Wronskian for $N$ ``seed'' functions  $\psi_{2s-1}=\cosh{\theta_{2s-1}}$, $\psi_{2s}=\sinh{\theta_{2s}}$ for integer $s \ge 1$, where the phases are $\theta_j = k_j(x-V_j t-x_j)$, $j=1,2,...,N$. The eigenvalues $k_j^2$ of the associated scattering problem specify the soliton amplitudes $A_j=2k_j^2$ and velocities $V_j=4k_j^2$, while the constants $x_j$ are responsible for the respective positions of solitons at a given time. The solution $u_N(x,t)$ is always positive \cite{Gardneretal1974}.

The averaged quantities in the relation (\ref{Variance}) can be calculated explicitly \cite{Pelinovskyetal2013,DutykhPelinovsky2014} for the $N$-soliton solution (\ref{NKdVSoliton}) in the asymptotic limit,
\begin{align} \label{Asymptotics}
	u_N(x,t) \underset{t \to \pm \infty}{\longrightarrow}  \sum_{j=1}^{N} A_j \text{sech}^2 \left(k_j(x-V_j t-x_j^{\pm}) \right),
\end{align}
where $x_j^{\pm}$ are reference locations of the solitons in the limits $t \to \pm \infty$. They read: 
$\overline{u} = 4 \rho \langle k \rangle$ and $\overline{u^2} = \frac{16}{3} \rho \langle k^3 \rangle$, where the angle brackets denote averaging in the spectral domain. According to these relations, the critical soliton density reads \cite{PelinovskyShurgalina2017}
\begin{align} \label{rho_cr_Def}
	\rho_{cr} = \frac{\langle k^3 \rangle} {3 \langle k \rangle^2}.
\end{align}
Strictly speaking, solitons are not compact solutions, and hence the scale $L$ implied by the quantity $\rho$ is not well-defined in this illustrative example. Remarkably, the same expression (\ref{rho_cr_Def}) was obtained for interacting solitons in the thermodynamic limit state in Ref.~\onlinecite{El2016}, where the soliton density characteristic was introduced in a rigorous manner.

%=======================================================
%\emph{Formulation of the problem.---}
\section{Formulation of the problem} \label{sec:Formulation}
\begin{figure}[htp]
	\begin{centering}
		\includegraphics[width=9cm]{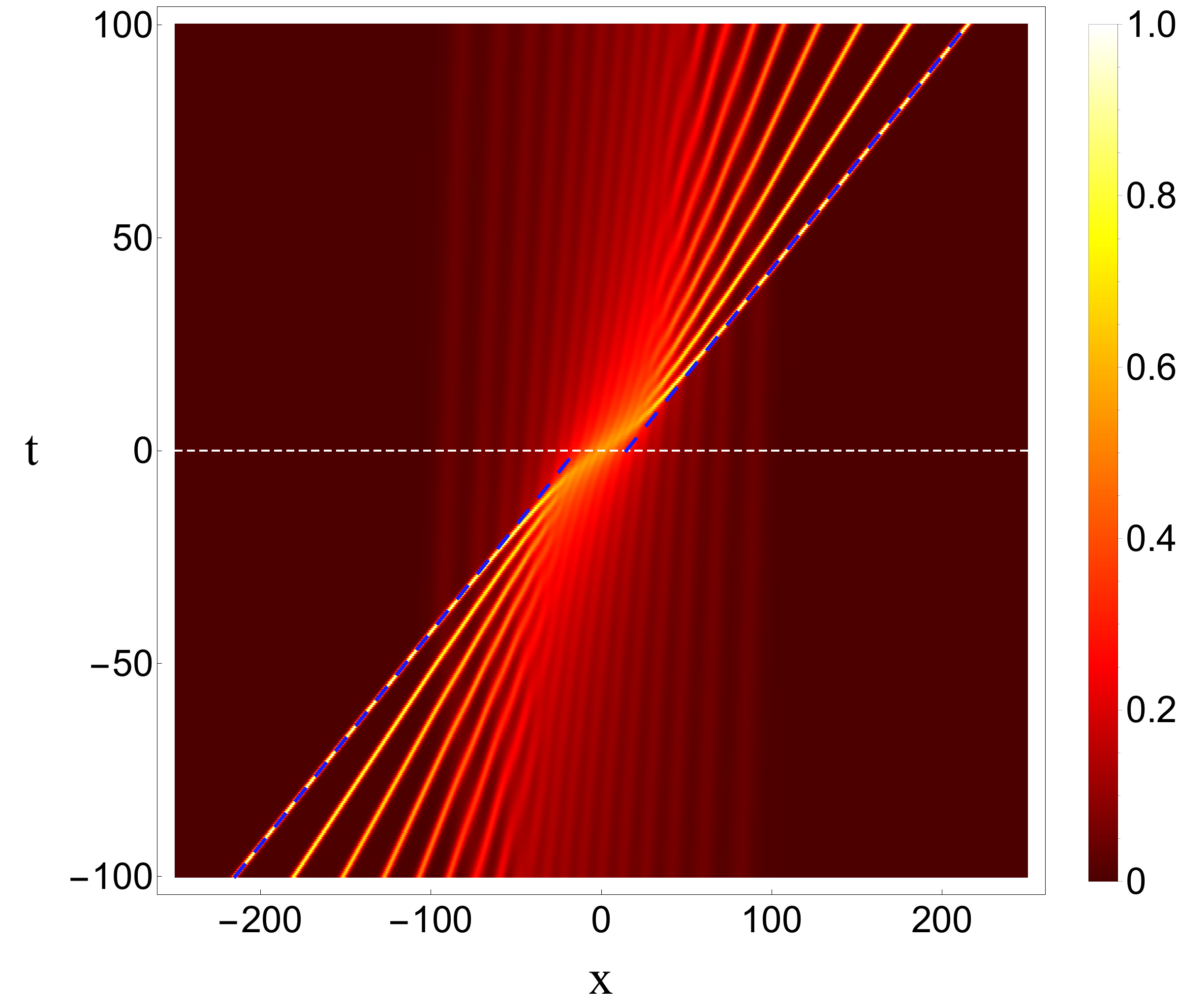}
		\caption{Interaction of $N=20$ KdV solitons with $d=1.2$. The blue dashed lines show the band of the spatial size $l_{foc}=\Delta x_1$
			\label{fig:SolitonCollisionTopVew}}
	\end{centering}
\end{figure}

%=======================================================
%
In the existing literature, most of researchers consider the limit of a small density which corresponds to the ``rarefied'' gas, when solitons interact in some sense weakly, and when approximate approaches can be efficient.
In the present work we are not confined by the assumption of a small density of solitons. Quite the opposite, we consider the extreme soliton superposition, which occurs in synchronous soliton collisions, when the reference locations of all the solitons at $t=0$ coincide with the coordinate origin. This property can be formalized through the following symmetry condition, $u_N(-x,-t)=u_N(x,t)$; it is fulfilled when all the parameters $x_j$ in (\ref{NKdVSoliton}) are put equal to zero: $x_j=0$, $j=1,...,N$. An example of such an interaction is shown in  Fig.~\ref{fig:SolitonCollisionTopVew}.

In what follows, the soliton amplitudes are set decaying exponentially, so that they form a geometric series with the ratio $d>1$:
\begin{align} \label{Amplitude_Law}
 A_j = \frac{1}{d^{j-1}}, \quad j=1,...,N. 
\end{align}
Within the KdV framework this choice is in fact rather generic as it corresponds to the distribution of eigenvalues of the scattering problem (represented by the stationary Schr\"{o}dinger equation) for a parabolic potential, which may serve as a first approximation to any hump-like disturbance.   

In the following Secs.~\ref{sec:States}--\ref{sec:Sech2} the statistical moments of synchronously interacting solitons are examined within the KdV framework. The applicability of the obtained results to a much broader class of equations is discussed in Sec.~\ref{sec:Generality}. We relate the fields of focused solitons to the states of the critical density in Sec.~\ref{sec:Localization}. Some final conclusions and perspectives are given in Sec.~\ref{sec:Conclusion}.

%=======================================================
%\emph{Quasi-stationary states of interacting solitons.---}
\section{Quasi-stationary states of interacting solitons} \label{sec:States}
%=======================================================
%
The following integrals will be used to measure the statistical moments, similar to Ref.~\onlinecite{Pelinovskyetal2013}:
\begin{align} \label{StatMoments Def}
	\mu_n (t) = \int_{-\infty}^{+\infty} {u^n(x,t) dx}, \quad n =1,2, ... . 
\end{align}
The choice of the soliton amplitude series (\ref{Amplitude_Law}) ensures that the asymptotic values $\mu_n(t \to \pm \infty)$ remain finite for any $n$ when $N \to \infty$. 

\begin{figure*}[htp]
	\begin{centering}
		\includegraphics[width=14cm]{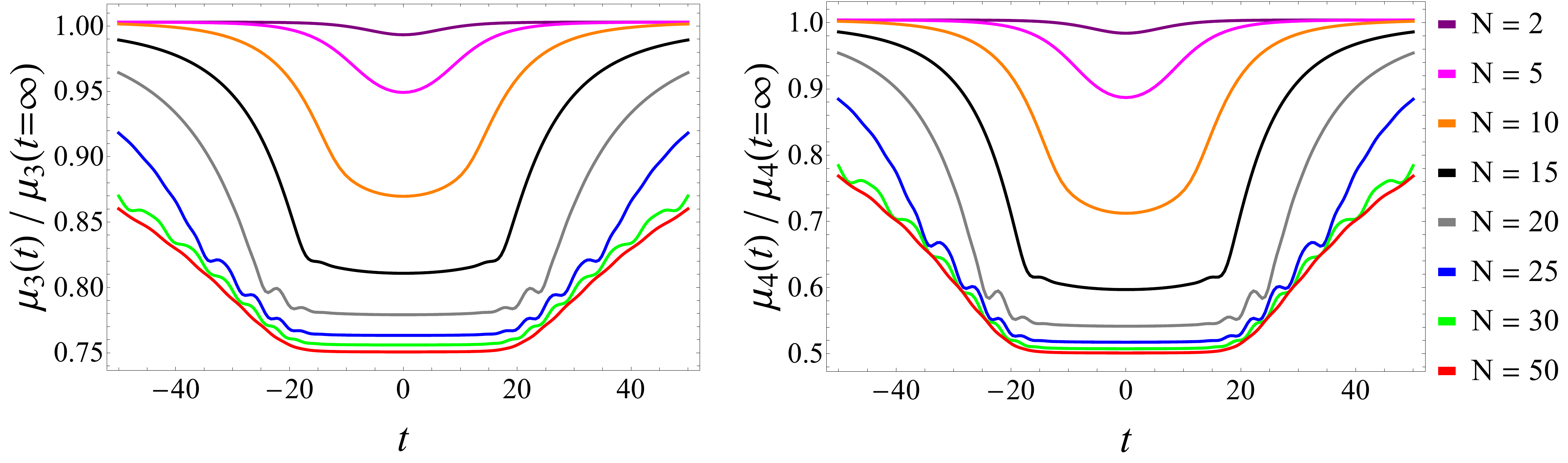}
		\includegraphics[width=14cm]{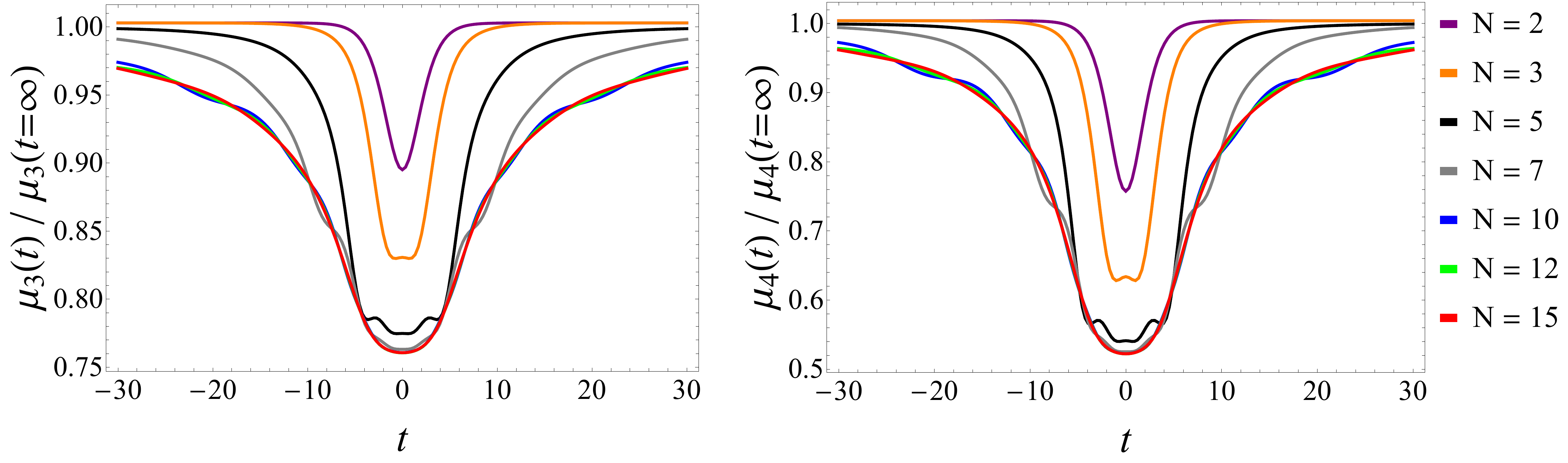}
		\caption{Evolution of relative statistical moments $\mu_3(t)/\mu_3(\infty)$ (left) and $\mu_4(t)/\mu_4(\infty)$ (right) for $N$-soliton solutions with $d = 1.1$ (top) and $d=1.6$ (bottom)}
		\label{fig:MomentsEvolution}
	\end{centering}
\end{figure*}

The use of an ultra-high-precision procedure made it possible to compute the exact $N$-soliton solutions (\ref{NKdVSoliton}) and to calculate the statistical moments (\ref{StatMoments Def}) for them, when $N$ is large, see details in Ref.~\onlinecite{TarasovaSlunyaev2022}.
%The inaccuracy of the first ten conservation laws for the solutions is within a fraction of per cent when $N \lesssim 50$, see details in \cite{TarasovaSlunyaev2022}. 
%
We found that the moments $\mu_3(t)$ and $\mu_4(t)$, which characterize the wave field asymmetry and kurtosis respectively, decrease in all situations when KdV solitons interact. This observation is in line with the limit of a two-soliton interaction \cite{Pelinovskyetal2013} and with the direct numerical simulations of rarefied soliton gases (see e.g. Refs.~\onlinecite{PelinovskyShurgalina2017,Didenkulova2019}). Representative examples of the evolution of $\mu_3(t)$ and $\mu_4(t)$ are given in Fig.~\ref{fig:MomentsEvolution} for two choices of the parameter $d$ and different numbers of interacting solitons $N$. 
It follows from the figures that the curves converge to some limiting ones when $N \gg 1$. For the values of $d$ closer to $1$ the convergence occurs at larger numbers $N$.

When $d$ is sufficiently close to $1$, and $N$ is large, the statistical moments remain approximately constant within long time spans, when the solitons are located most densely. This effect is readily seen in the upper panels in Fig.~\ref{fig:MomentsEvolution} ($d=1.1$) for $N \gtrsim 10$, but vanishes for smaller numbers $N$ or for significantly larger values of $d$ (see the bottom panel in Fig.~\ref{fig:MomentsEvolution} for $d=1.6$). When $d$ is large, the soliton amplitudes in the sequence (\ref{Amplitude_Law}) decay quickly, so that apparently only a few of the largest solitons play a role.
Obviously, the observation of ``plateaux'' in the dependencies $\mu_3(t)$, $\mu_4(t)$ should mean some degeneracy of the soliton state, which we explore below.

%=======================================================
%\emph{Virtual ``smoothing'' of the wave field in the focusing area.---}
\section{Smoothing of the wave field in the focusing area} \label{sec:Smoothing}
%=======================================================
%
It is well-known that the integrability property of the KdV equation is deeply related to the existence of an infinite series of conserved quantities $I_m$,
\begin{align} \label{I_m}
	\frac{d}{dt} I_m = 0, \quad I_{m} = \int_{-\infty}^{+\infty} T_{m}(u) dx, \quad m = 1,2, ... .
\end{align}
Hereafter we will use the densities $T_m$ in the form given in Ref.~\onlinecite{Miura1968}. The two first statistical moments $\mu_1$ and $\mu_2$ are proportional to the integrals $I_1=6\int_{-\infty}^{\infty}{u dx}$ and $I_2=18 \int_{-\infty}^{\infty}{u^2 dx}$, and thus do not change in time, while the higher-order statistical moments may vary. 

The third conservation integral may be presented in the form
\begin{align} \label{I3}
	I_3 = 72 \left( \mu_3 -\chi_3 \right), 
	\quad
	\chi_3(t)=\frac{1}{2} \int_{-\infty}^{\infty}{  (u_x)^2  dx} . 
\end{align}
With the use of the asymptotic solution (\ref{Asymptotics}), the terms in (\ref{I3}) can be explicitly computed for $t\to \pm \infty$,
\begin{align} \label{I3I4_Values}
	\mu_3(\infty) = \frac{128 N \langle k^5 \rangle}{15},
	\quad
	\chi_3(\infty) = \frac{1}{4} \mu_3(\infty),
\end{align}
and then
\begin{align} 
	\frac{\mu_3(t)}{\mu_3(\infty)} = \frac{3}{4} + \frac{\chi_3(t)}{\mu_3(\infty)}. \label{M3_Value}
\end{align}
It was noted in Ref.~\onlinecite{Pelinovskyetal2013} that since $\mu_3$ and $\chi_3$ are strictly positive quantities,  the decrease of $\mu_3$ in the course of the collision leads to the decrease of the second integral, $\chi_3$. This should correspond to ``smoothing'' of the solution in some sense. According to (\ref{M3_Value}), the minimal possible value of $\mu_3$, $\mu_{3,min}/\mu_3(\infty) = 3/4$ is achieved when $\chi_3$ totally vanishes.  Note that this limit well agrees with the minimum values of $\mu_3$ observed in Fig.~\ref{fig:MomentsEvolution} (left column).

% At the same time, the anticipated wave ``smoothing'' is hardly visible by an eye in the illustration in Fig.~\ref{fig:FocussedTrain} where the $N$-soliton solution at the focusing instant is shown against the shapes of isolated solitons. 

In order to elucidate the nature of the ``smoothing'' of the wave field, it is instructive to introduce the similarity parameter $\epsilon(t)$ (the so-called Ursell number), which estimates the ratio of the nonlinear term versus the term of the wave dispersion in the evolution equation (\ref{KdV}) for a given solution $u(x,t)$ at a given instant of time. 
%Assuming that a solution has a characteristic amplitude $A$ and spatial length $L$, then $\epsilon \sim AL^2$. 
Then the conservation integrals (\ref{I_m}) may be presented in the form 
\begin{align} 
	I_m = \frac{6^m}{m} \mu_m \left(1 + O(\epsilon^{-1}) \right), \label{StructureOfIntegrals}
\end{align}
see Ref.~\onlinecite{Karpman1975}. In these notations, the ``smoothing'' should correspond to the occurrence of very large values of $\epsilon$ (in other  words, the small-dispersion limit) at $t\approx 0$. 
From this, we obtain the estimate for $\mu_m(0)$, and then the general formula for any integer $n \ge 1$ assuming $\epsilon \gg 1$:
\begin{align} 
	\frac{\mu_n(0)}{\mu_n(\infty)} =   \frac{n I_n}{6^n \mu_n(\infty)} \left( 1 + O(\epsilon^{-1}) \right) = \frac{2n}{2^n} \left(1 + O(\epsilon^{-1}) \right). \label{Mm_Value}
\end{align}
To obtain the final expression in (\ref{Mm_Value}), the integrals $I_n$ and $\mu_n(\infty)$ should be calculated for the asymptotic solution  (\ref{Asymptotics}):  
\begin{align}\label{I_n_Value}
	I_n= 24^{n}
	\frac{[(n-1)!]^2}{(2n-1)!} N \langle k_j^{2n-1} \rangle , \\	
	\label{Mm_Inf_Value}
	\mu_n (\infty)=
	2^{3n-1} \frac{[(n-1)!]^2}{(2n-1)!} N \langle k_j^{2n-1} \rangle. 	
\end{align}
The integrals $I_n$ for a single soliton may be found in Ref.~\onlinecite{Karpman1975}.

\begin{figure}[htp]
	\begin{centering}
		\includegraphics[width=9cm]{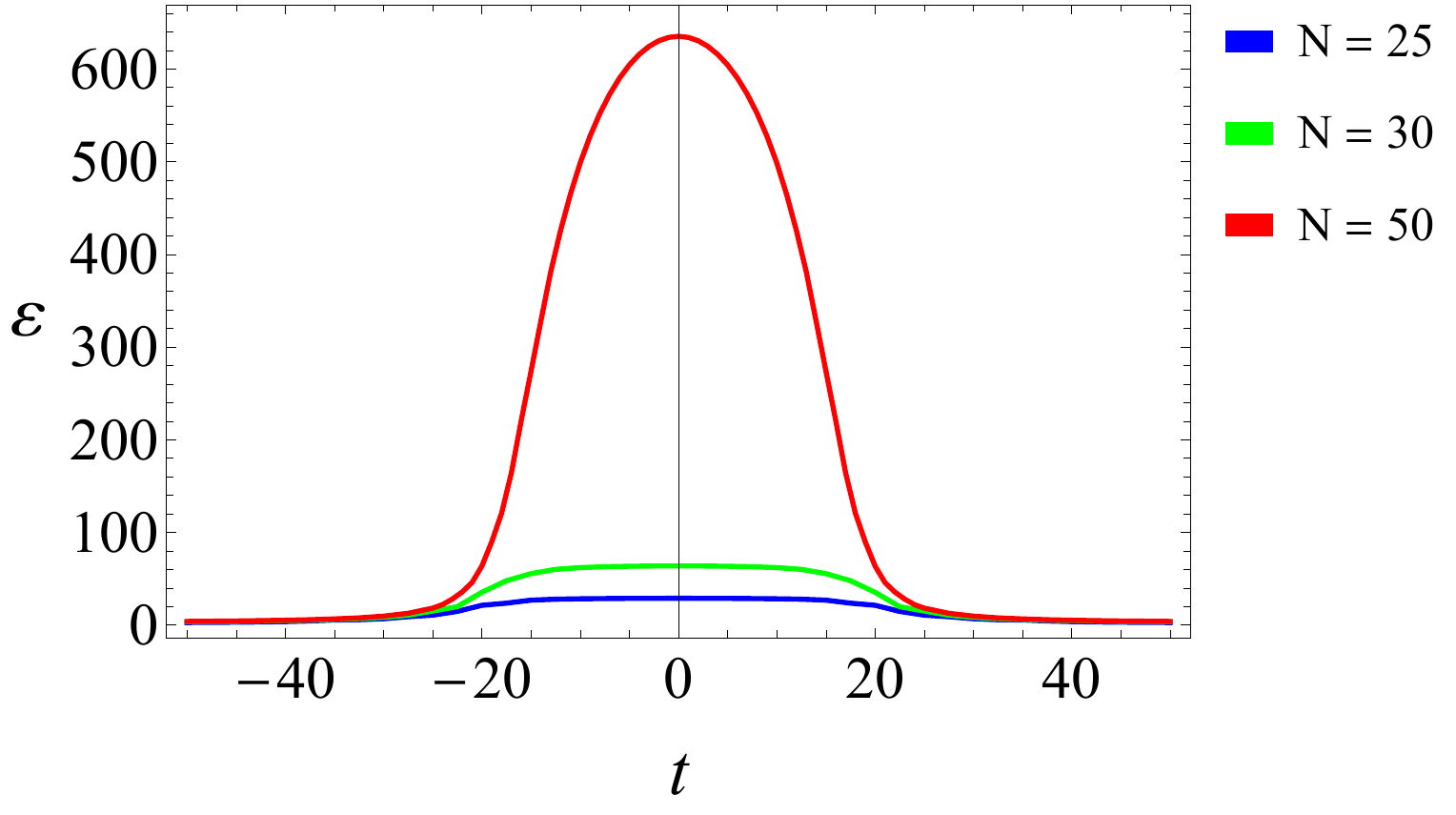}
		\caption{Evolution of the Ursell number $\epsilon$ during synchronous interactions of KdV solitons with $d=1.1$ for different $N$
			\label{fig:UrsellParameterKdV}}
	\end{centering}
\end{figure}

In Fig.~\ref{fig:UrsellParameterKdV} the evolution of the Ursell number for the case $d=1.1$ is shown for different $N$. The Ursell number is calculated using the integral estimator $\epsilon = \int_{-\infty}^{\infty}{u^3 dx} / \int_{-\infty}^{\infty}{(u_x)^2 dx} = \mu_3 / {(2\chi_3)}$. This quantity has the value $\epsilon = 2$ for free solitons of any amplitude. One can see from the figure that indeed, the values of $\epsilon$ become very large when the solitons interact. They further grow if $N$ increases. A large number of solitons with close velocities leads to a long time of interaction, what explains the appearance of plateaux in the dependencies $\mu_3(t)$ and $\mu_4(t)$ in Fig.~\ref{fig:MomentsEvolution} (top). For $d=1.6$ the parameter $\epsilon$ increases considerably as well, but due to a shorter time of interaction a quasi-stationary behavior of the statistical moments is not observed (Fig.~\ref{fig:MomentsEvolution}, bottom).

One can easily check that the relation (\ref{Mm_Value}) aligns with constant moments $\mu_1$ and $\mu_2$, and provides the limiting values $3/4$ and $1/2$ for the drops of the third and fourth moments, which agree with the numerical solutions shown in Fig.~\ref{fig:MomentsEvolution}. It follows from the figure, that the estimates (\ref{Mm_Value}) are applicable to long time spans when the solitons focus, if $d$ is close to $1$ and $N\gg1$. According to (\ref{Mm_Value}), statistical moments of any order decrease when KdV solitons interact, and this effect becomes stronger for higher moments.
We have checked numerically up to the order $n=7$ that the estimate (\ref{Mm_Value}) is remarkably accurate, see Table~\ref{tab:Moments}.

\begin{figure}[htp]
	\begin{centering}
		\includegraphics[width=8.5cm]{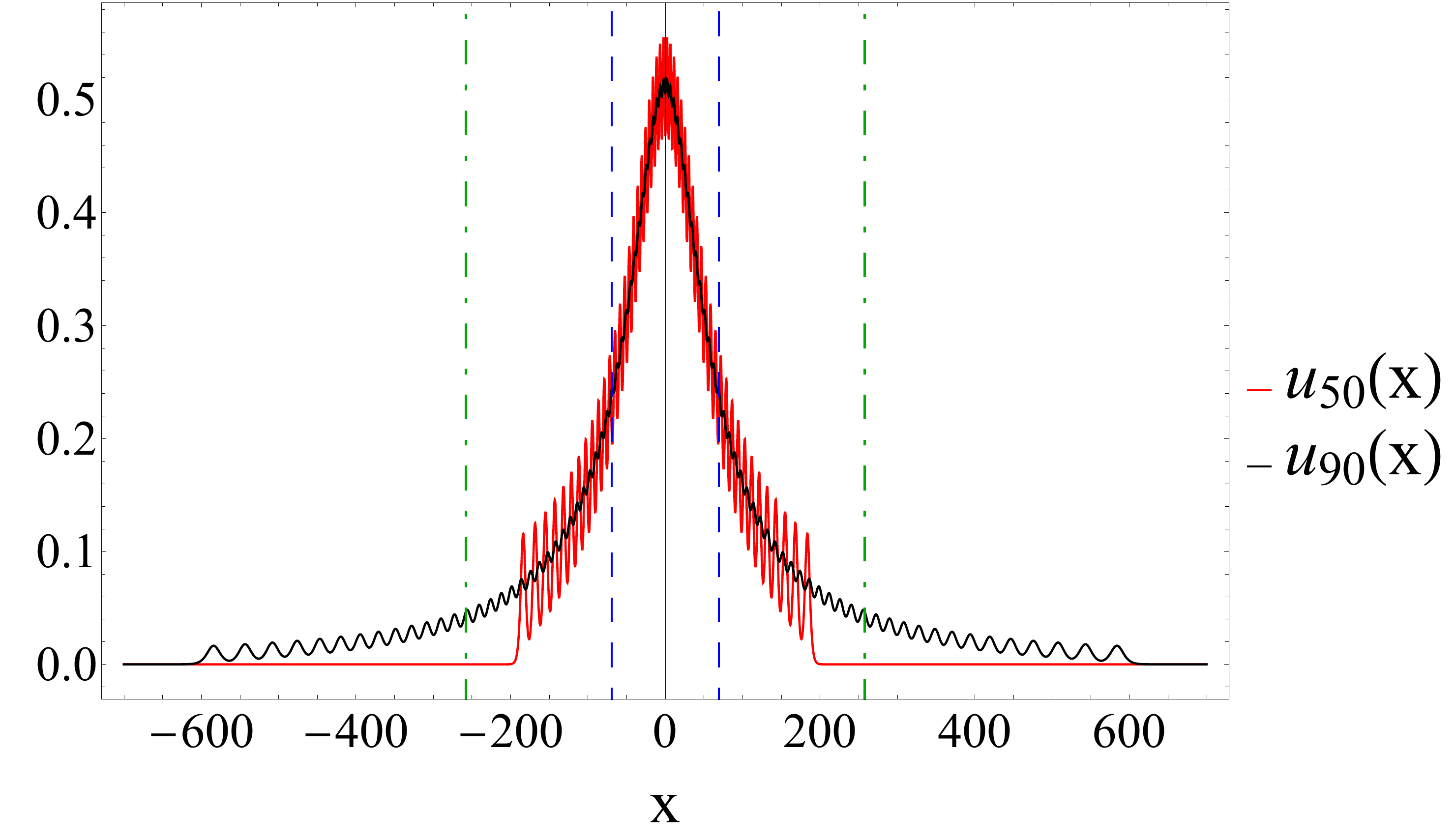}
		\caption{$N$-soliton solutions $u_N$ for $d = 1.05$ and $N=50$, $90$ at $t=0$. The dashed blue lines $x = \pm l_{foc}/2$ and dashdot green lines $x = \pm l_{cr}/2$ are shown for the limit of infinite number of solitons $N \to \infty$}
		\label{fig:FocussedTrain}
	\end{centering}
\end{figure}

Examples of the numerical solutions $u_N$ for $d$ close to $1$ and different finite number of solitons $N$ are shown in Fig.~\ref{fig:FocussedTrain} for the instant $t=0$. When $N$ grows, the wave smoothing becomes apparent through a decrease in local maxima and an increase in the amplitude of the solution within the valleys between the humps.    
Accounting for a greater number of solitons in the series (\ref{Amplitude_Law}) leads to further smoothing of the solution and modification of its peripheral part.
Presumably, in the limit of a large number $N$ of solitons with amplitudes distributed according to the geometric progression they focus into a visually one-hump profile for any $d> 1$. A similar effect in reconstructing a box potential by solitons of the nonlinear Schr\"{o}dinger equation was found in Ref.~\onlinecite{Gelashetal2021}.

%\red{Presumably, in the case of $d \approx 1$ and a large number $N$ of solitons with a power-law amplitude distribution, a great Ursell number corresponds to the focusing of solitons into a visually one-hump profile with heavy tails, see Fig.~\ref{fig:FocussedTrain}. }

\begin{table}
	\caption{\label{tab:Moments}Deviation of the numerical solutions of the KdV and the mKdV equations ($d=1.1$ and $N=50$) from the theoretical minimum relative values of the $n$-th statistical moments $\mu_n(0)/\mu_n(\infty)$}
	\begin{ruledtabular}
		\begin{tabular}{cccc}
			$n$ & analytical estimate &  deviation, KdV & deviation, mKdV \\ \hline
			$3$ & $3/4$ & 0.08\% & 0.36\%\\
			$4$ & $1/2$ & 0.22\% & 0.58\%\\
			$5$ & $5/16$  & 0.42\% & 0.86\% \\
			$6$ & $3/16$ & 0.72 \% & 1.17\%\\
			$7$ & $7/64$ & 1.1\% & 1.67\%\\
		\end{tabular}
	\end{ruledtabular}
\end{table}

%=======================================================
%\emph{Representative analytical solution.---}
\section{Representative analytical solution} \label{sec:Sech2}
%=======================================================
%
There is a well-known particular shape of the initial perturbation which gives purely discrete spectrum of the associated scattering problem for the KdV equation \cite{Lamb1980}:
\begin{align} \label{SechShape}
	u(x,0) = N (N+1) \text{sech}^2{x}.
\end{align}
Its evolution leads to the emergence of $N$ solitons with zero dispersive tail. The spectral parameters are given by the series
\begin{align} \label{EigenValues}
	k_j = N-j+1, \quad j = 1, ..., N,
\end{align}
which yields the following relation between soliton amplitudes,
\begin{align} \label{AmplitudeRatioSech2}
	\frac{A_j}{A_{j+1}} = \left( 1+ \frac{1}{N-j} \right)^2	\underset{j \ll N}{\longrightarrow} 1+ \frac{2}{N} > 1.
\end{align}
Thus, it is close to the sequence (\ref{Amplitude_Law}) with the ratio $d = 1+2/N$, when $N$ is large and the soliton sequence numbers $j$ are small (i.e., for the tallest solitons). 

For the initial-value problem specified by (\ref{SechShape}), the ratio for an arbitrary $n$-th statistical moment can be evaluated in a few steps using the asymptotic solution (\ref{Asymptotics}) and assuming that the number of solitons is large:
\begin{align}  \label{FormulaForMoments}
	\frac{\mu_n(0)}{\mu_n(\infty)} = \frac{N^n \left( N+1 \right)^n}{2^n \sum_{j=1}^{N}{k_j^{2n-1}}}
	\underset{N \gg 1}{\longrightarrow} \nonumber \\
	\frac{N^{2n}}{2^n \sum_{j=1}^{N}{\left( N-j+1 \right)^{2n-1}}} \underset{N \gg 1}{\longrightarrow}
	\frac{2n}{2^n}.
\end{align}
Thus, we obtain exactly the same result (\ref{Mm_Value}) provided by the qualitative description in the previous section.   

%=======================================================
%\emph{Broader generality of the result.---}
\section{Broader generality of the result} \label{sec:Generality}
%=======================================================
%
Since the formulae (\ref{I_n_Value}) and (\ref{Mm_Inf_Value}) are general, the obtained result (\ref{Mm_Value}) is valid for any distribution of the soliton amplitudes, as soon as the solution does not change the sign and the similarity parameter $\epsilon$ is large; the considered example of a $\text{sech}^2$ perturbation (\ref{SechShape}) supports this conclusion.
This dynamic scenario may be considered as a particular realization of a dense soliton gas.
The discovered statistical property of ensembles of a large number of unipolar solitons actually has a much broader application beyond the KdV framework.  Below we give a few examples of this fact.

First, the relation (\ref{Mm_Value}) is applicable to all members of the \textit{hierarchy of integrable KdV equations} with the scattering problem represented by the stationary Schr\"odinger equation. Indeed, in this case the $N$-soliton solutions may be constructed using the same Darboux transformation \cite{MatveevSalle1991}. The difference will be in the time-dependence of solitons only (i.e., other expressions for the soliton velocity $V_j(k_j)$), but this difference vanishes at the focusing moment $t=0$ and has no effect on the statistical moments when the solitons separate at $t \to \pm \infty$. Consequently, the moments $\mu_n(0)$ and $\mu_n(\infty)$ will have exactly the same values as before. 
The obtained results obviously can be extended to higher dimensions, such as the \textit{Kadomtsev -- Petviashvili equation}.
%Below, we consider briefly two extra examples.     

The focusing \textit{modified Korteweg -- de Vries} (mKdV) equation 
\begin{align}\label{mKdV}
	w_t + 6 w^2 w_{x} + w_{xxx} = 0
\end{align}
has a different from the classic KdV associated scattering problem, but as other integrable systems, possesses an infinite number of conserved quantities $I_m^{(mKdV)}$, $m=1,2,...$. Soliton solutions of the mKdV equation have two branches of solitons on a zero background which differ in a sign (polarity).

\begin{figure}[htp]
	\begin{centering}
		\includegraphics[width=9cm]{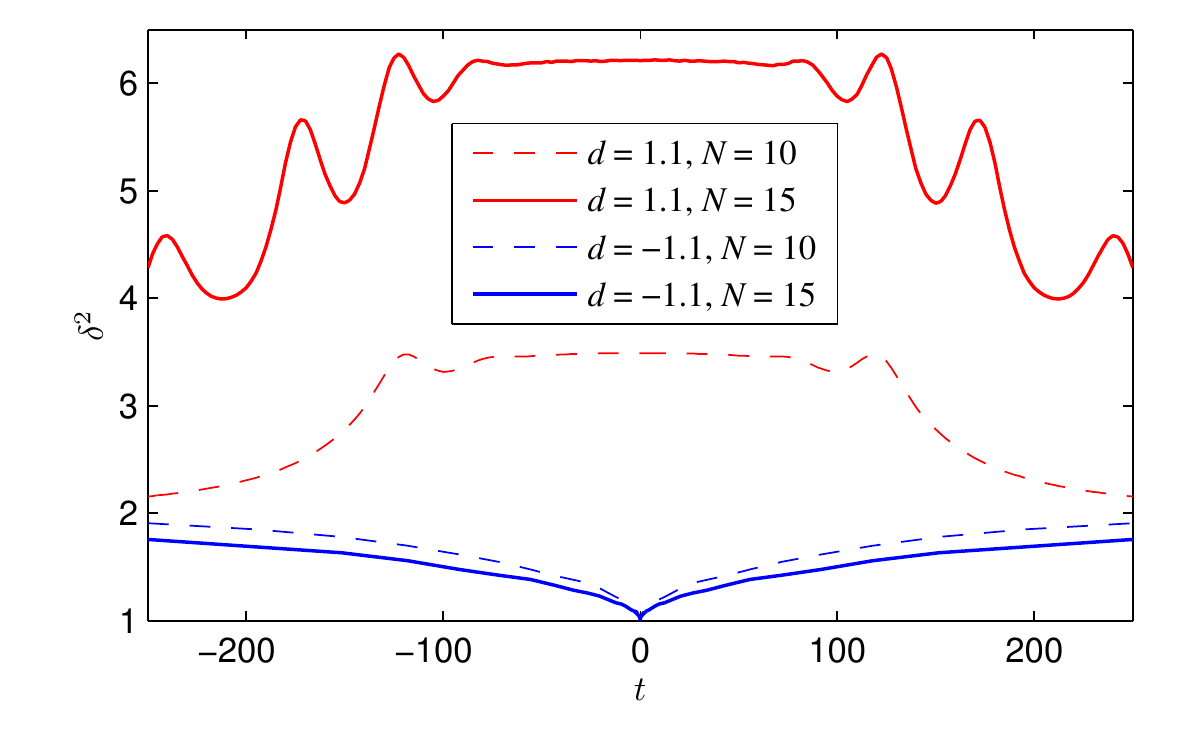}
		\caption{Evolution of the parameter $\delta^2$ during synchronous interactions of mKdV solitons with the same signs ($d=1.1$) and alternating signs ($d=-1.1$)
			\label{fig:SimilarityParameterMKdV}}
	\end{centering}
\end{figure}

If the similarity parameter of the mKdV equation $\delta$, which may be estimated as  $\delta^2 = \int_{-\infty}^{\infty}{w^4 dx}/ \int_{-\infty}^{\infty}{w_x^2 dx}$ becomes large (i.e., the small-dispersion regime takes place), then to the leading order the conservation integrals are proportional to the statistical moments, $I_m^{(mKdV)} \propto \mu_{2(m-1)}^{(mKdV)} \left( 1 + O(\delta^{-2}) \right)$, $\mu_{n}^{(mKdV)} = \int_{-\infty}^{\infty}{w^{n}dx}$. 
The evolution of the parameter $\delta$ in two examples of interaction of mKdV solitons of the same sign is shown in Fig.~\ref{fig:SimilarityParameterMKdV} with red curves. 
A similar approach based on the Darboux transform is used to construct $N$-soliton solutions of the mKdV equation, see Ref.~\onlinecite{Slunyaev2019}. The soliton amplitudes are specified according to the same law (\ref{Amplitude_Law}) as in the case of the KdV equation. During interaction, the parameter greatly exceeds the value $\delta^2=2$ which corresponds to free soliton solutions,  if $d$ is close to $1$ and $N$ is sufficiently large.
%, using an ultra-high-precision procedure.}

Therefore, when the nonlinear term in the equation dominates over the term of dispersion, one can straightforwardly estimate the relative decrease of the statistical moments with even orders $n$. Remarkably, these estimates exactly coincide with the ones for the KdV equation (\ref{Mm_Value}). Even more, we have checked numerically that odd orders of the statistical moments satisfy the relation (\ref{Mm_Value}) too, 
\begin{align}\label{Mm_mKdV}
	\frac{\mu^{(mKdV)}_{n}(0)}{\mu_n^{(mKdV)}(\infty)}=\frac{2n}{2^n} \left( 1 + O(\delta^{-2}) \right), \quad n=1,2,.... 
\end{align} 
Note the perfect agreement between numerical results and the analytical estimates in Table~\ref{tab:Moments}. 

In fact, equations (\ref{KdV}) and (\ref{mKdV}) are related by the complex Miura transformation
\begin{align}\label{Miura}
	q(x,t)=w^2 - i w_x,
\end{align}
which maps solutions of the mKdV equation $w(x,t)$ to complex-valued solutions $q(x,t)$ of the \textit{complex Korteweg -- de Vries} (cKdV) equation on $q(x,t)$ which has the form identical to (\ref{KdV}) (see e.g. Ref.~\onlinecite{Sunetal2014}). 

The balance of terms in the Miura transformation is controlled by the similarity parameter $\delta$ of the mKdV equation, which quantifies the contribution of the nonlinear term in (\ref{Miura}) with respect to the imaginary term of dispersion, so that the real part effectively dominates at the focal point $t \approx 0$. That provides an estimate $\int_{-\infty}^{\infty}{q^n(x,t\approx 0) dx} \approx \int_{-\infty}^{\infty}{w^{2n}(x,t\approx 0) dx}$. At asymptotically large times $|t| \gg 1$ solitons of the mKdV equation   $w_j(x,t)= 2\lambda_j \text{sech}{\left[ 2\lambda_j (x-4 \lambda_j^2 t) \right]}$ are mapped to $q_j(x,t)=2 \lambda_j^2 \text{sech}^2{\left[ \lambda_j(x-4 \lambda_j^2 t) - i \pi/4 \right]}$, where real $\lambda_j$ is now the spectral parameter of the scattering problem for the mKdV equation. The expression for $q_j(x,t)$ coincides with the single soliton solution for the KdV equation (see (\ref{Asymptotics}) putting $k_j=\lambda_j$) except for the imaginary phase shift. The imaginary phase shift does not influence the integrals $\int_{-\infty}^{\infty}{q^n dx}$, they remain real-valued. Therefore, one can straightforwardly obtain that $\int_{-\infty}^{\infty}{q^n(x,t \to \pm \infty) dx} = 2^{1-n} \int_{-\infty}^{\infty}{w^{2n}(x,t \to \pm \infty) dx}$.
As a result, the statistical moments for the solution $q(x,t)$ of the complex KdV equation, $\mu_{n}^{(cKdV)} = \int_{-\infty}^{\infty}{q^ndx}$, may be also expressed through the ratio (\ref{Mm_mKdV}):
\begin{align}\label{ComplexKdVMoments}
	\frac{\mu_n^{(cKdV)}(0)}{\mu_n^{(cKdV)}(\infty)} \approx 2^{n-1} \frac{\mu_{2n}^{(mKdV)}(0)}{\mu_{2n}^{(mKdV)}(\infty)} \approx \frac{2n}{2^n}, \quad n=1,2,... . 
\end{align}

%=======================================================
%\emph{Localization size of the focussed soliton train.---}
\section{Localization size of the focused soliton train} \label{sec:Localization}
%=======================================================
%
During the intervals of almost constant reduced statistical moments the interacting KdV solitons are located most closely and are characterized by large Ursell numbers. It seems reasonable to associate these extreme quasi-stationary states with the situation of the critical soliton density.

Importantly, the focused solitons represent essentially non-uniform in space wave fields, thus the adopted formulas, which relate the averaging in the physical space to the averaging in the spectral domain and eventually lead to the formula (\ref{rho_cr_Def}), become questionable. 
One can straightforwardly check that the definition (\ref{rho_cr_Def}) for the soliton amplitudes taken in the form of a geometric series (\ref{Amplitude_Law}) yield the values proportional to $N$, hence the quantity $\rho_{cr}$ grows infinitely if $N \to \infty$.
Bearing this in mind, let us consider the quantity $l_{cr}=N/\rho_{cr}$ which has the meaning of some characteristic size, assuming that $\rho_{cr}$ is still determined by (\ref{rho_cr_Def}). 
For the exponential distribution of the soliton amplitudes (\ref{Amplitude_Law}) the values of $l_{cr}$ are finite:
\begin{align} \label{length_cr_KdV}
	l_{cr} \underset{\substack{N \gg 1 \\ d \to 1+0}}{\longrightarrow} \frac{18 \sqrt{2}}{d-1}. 
\end{align}

To estimate the actual size of the spatial domain $l_{foc}$ occupied by the solution at $t \approx 0$, we consider the coordinate shift which solitons experience when pass through all the other solitons in the train (see Fig.~\ref{fig:SolitonCollisionTopVew}). The shift for the $p$-th soliton, $\Delta x_p = |x_p^{+} - x_p^{-}|$, is given by the classic formula
\begin{align} \label{DeltaX}
	\Delta x_p = \frac{1}{k_p} \sum_{\substack{j=1 \\ j\neq p}}^{N}{\Delta x_{p,j}}, \quad
	\Delta x_{p,j} =  \ln{\left|  \frac{k_p+k_j}{k_p-k_j} \right|}.
\end{align}
Due to the factor $k_p^{-1}$, the shift of small-amplitude solitons is not bounded from above when $k_p \to 0$. On the other hand, the shift of the largest soliton $\Delta x_1$ is the least affected by an addition of small-amplitude solitons. Therefore we represent the characteristic size of the focused solution through the shift of the largest soliton, $l_{foc}=\Delta x_1$. The blue dashed lines in  Fig.~\ref{fig:SolitonCollisionTopVew} correspond to the plots $(- \Delta x_1/2 + V_1 t,t)$ for $t<0$ and $(\Delta x_1/2 + V_1 t,t)$ for $t>0$. They follow the paths of the largest solitons when $|t|>>1$ and confine the interval $-l_{foc}/2 \le x \le l_{foc}/2$ when cross the axis $t=0$.
For the soliton amplitudes distributed according to (\ref{Amplitude_Law}), this quantity can be calculated assuming $N$ is large and $d$ is close to $1$,
\begin{align} \label{DeltaX_Solution}
	l_{foc} = \Delta x_1 \underset{\substack{N \gg 1 \\ d \to 1+0}}{\longrightarrow} \frac{\pi^2}{\sqrt{2} \ln{d}}
	\underset{\substack{N \gg 1 \\ d \to 1+0}}{\longrightarrow} \frac{\pi^2}{\sqrt{2} (d-1)}.
\end{align}
The estimated size of the focused wave has the same dependence on the governing parameter $d$ as the critical size (\ref{length_cr_KdV}), with the proportionality coefficient $l_{cr}/l_{foc} \approx 3.65$.  

For the shape (\ref{SechShape}) the characteristic size $l_{foc}$ of the focused solution can be estimated  using (\ref{DeltaX}) as $l_{foc} = \Delta x_1 \approx 2 \ln{2}$.  One can also calculate the critical width $l_{cr}=N/\rho_{cr}$ for this example directly using the formula for the critical density (\ref{rho_cr_Def}) and the distribution of soliton amplitudes according to (\ref{EigenValues}), $l_{cr}=3$; so for this case $l_{cr}/l_{foc} \approx 2.16$.

%\red{The characteristic size $l_{foc}$ of the focused solution can also be estimated for the shape (\ref{SechShape}) using  (\ref{DeltaX}), $l_{foc} = \Delta x_1 \approx 2 \ln{2}$. Note that the visible size of (\ref{SechShape}) is not changed drastically with a considerable increase of $N$, hence the obtained constant value is rather a plausible estimate. One can also straightforwardly calculate the value $l_{cr}=N/\rho_{cr} = 3$, so $l_{cr}/l_{foc} \approx 2.16$}.

%\red{For the $N$-soliton solution shown in Fig.~\ref{fig:FocussedTrain} the characteristic sizes are $l_{cr} \approx 407.7$ and $\Delta x_1 \approx 125.5$. It should be mentioned that $l_{cr}$ is in good compliance with the visible spatial scale of the focused wave train.}

For the examples of multisoliton solutions $u_{50}$ and $u_{90}$ shown in Fig.~\ref{fig:FocussedTrain} the characteristic sizes are $l_{cr} \approx 263$, $\Delta x_1 \approx 104$ and $l_{cr} \approx 408$, $\Delta x_1 \approx 126$ correspondingly. 
It may be seen from the figure that the scales $l_{foc}$ and $l_{cr}$ are in good compliance with the visible size of the focused wave train.	
One may speculate that the extreme compression of solitons which occurs in the considered scenarios of soliton collisions corresponds to the states with the minimum allowed (critical) size of the soliton ensemble.  
The scale $l_{foc} = \Delta x_1$ underestimates the actual visible size of the focused solution, and can be used as the estimate from below.

%=======================================================
%\emph{Conclusion.---}
\section{Conclusion} \label{sec:Conclusion}
%=======================================================
%
In this work we present a general idea that dense ensembles of solitons of the same sign (say, KdV-type solitons) can be considered as the strongly-nonlinear / small-dispersion wave states, what allows to express the statistical moments in terms of the spectral parameters of the associated scattering problem. Synchronous collisions of many solitons with slowly decaying amplitudes have been considered as a particular case when the dense soliton state can occur. We made a qualitative assumption (without a rigorous justification) on the relation between the critical soliton density and the minimal size of the focused solitons, which shows a reasonable agreement.

A broad applicability of the employed concept to other integrable systems is anticipated, we have given several confirming examples. Since weakly non-integrable systems typically inherit the soliton dynamics with little alteration (see e.g. Ref.~\onlinecite{DutykhPelinovsky2014}), we may expect that the obtained results can be applied to non-integrable generalizations of the equations too. 

At the same time, the approach apparently does not allow one to estimate the statistical moments in the case of simultaneous collisions of solitons with different signs, when they can cause wave amplification, which is not limited from above with an increase in the number of solitons \cite{SlunyaevPelinovsky2016,Slunyaev2019}. 
It follows from the direct numerical calculations that in the course of collision of mKdV solitons with alternating signs, the similarity parameter associated with the integral ratio, $\delta^2 = \int_{-\infty}^{\infty}{w^4 dx}/ \int_{-\infty}^{\infty}{w_x^2 dx}$, decreases from the value $\delta^2=2$ which corresponds to free soliton solutions, but not much. It never falls below the value $\delta=1$, see the blue curves in Fig.~\ref{fig:SimilarityParameterMKdV}. Hence, a discrimination of terms which constitute the conservation integrals does not occur. Since phases of soliton solutions are not affected by their signs, the synchronous collision of solitons with alternating polarities formally leads to the creation of an equally dense soliton state, as in the case when solitons have the same signs. However, the statistical moments of the focusing solitons of different signs change rapidly, in contrast to the case of unipolar solitons.

In that way, a counterintuitive situation takes place, when interacting solitons of the same sign produce a strongly nonlinear state in terms of the similarity parameter, but which does not contain high waves. At the same time, the collision of solitons of different signs may cause extremely high wave amplitudes but is characterized by a relatively small ratio of nonlinearity versus dispersion.

\section*{Author's contributions}
All authors contributed equally to this work.

\begin{acknowledgments}
The authors are grateful to M.V. Pavlov and E.N. Pelinovsky for valuable stimulative discussions.
The work on Sec.~\ref{sec:Generality} was supported by the Foundation for the Advancement of Theoretical
Physics and Mathematics ``BASIS'' No~22-1-2-42. The remaining sections were supported by the RSF grant No~19-12-00253. The authors are grateful to the anonymous referee for valuable comments.
\end{acknowledgments}

\section*{Data Availability Statement}
Data available on request from the authors.

\bibliography{Solitons_v3}% Produces the bibliography via BibTeX.

\end{document}